\documentclass[aps,prd,twocolumn,showpacs,amsmath,amssymb]{revtex4-1}
\usepackage{amsmath}
\usepackage{graphicx}
\usepackage{subfigure}
\usepackage{epstopdf}
\usepackage{color}
\usepackage{multirow}
\usepackage{setspace}
\usepackage{overpic}
\usepackage{amssymb}
\usepackage[bookmarksnumbered, pdfstartview=FitH,colorlinks,urlcolor=blue, citecolor=blue,linkcolor=blue] {hyperref}
\usepackage{lineno}
\usepackage{bm}
\usepackage{rotating}
\usepackage{verbatim}
\usepackage[utf8]{inputenc}
\let\oldequation\equation
\let\oldendequation\endequation

\renewenvironment{equation}
  {\linenomathNonumbers\oldequation}
  {\oldendequation\endlinenomath}

\begin{document}


\title{\boldmath Search for the lepton number violation decay $\phi \to \pi^+ \pi^+ e^- e^-$ via $J/\psi\to \phi\eta$
}

\author{
M.~Ablikim$^{1}$, M.~N.~Achasov$^{13,b}$, P.~Adlarson$^{73}$, R.~Aliberti$^{34}$, A.~Amoroso$^{72A,72C}$, M.~R.~An$^{38}$, Q.~An$^{69,56}$, Y.~Bai$^{55}$, O.~Bakina$^{35}$, I.~Balossino$^{29A}$
% ...
\\
\vspace{0.2cm}
(BESIII Collaboration)\\
\vspace{0.2cm} {\it
$^{1}$ Institute of High Energy Physics, Beijing 100049, People's Republic of China\\
% ...
$^{13}$ G.I. Budker Institute of Nuclear Physics SB RAS (BINP), Novosibirsk 630090, Russia\\
% ...
$^{29}$ INFN Sezione di Ferrara, (A)INFN Sezione di Ferrara, I-44122, Ferrara, Italy; (B)University of Ferrara,  I-44122, Ferrara, Italy\\
% ...
$^{34}$ Johannes Gutenberg University of Mainz, Johann-Joachim-Becher-Weg 45, D-55099 Mainz, Germany\\
$^{35}$ Joint Institute for Nuclear Research, 141980 Dubna, Moscow region, Russia\\
% ...
$^{38}$ Liaoning Normal University, Dalian 116029, People's Republic of China\\
% ...
$^{55}$ Southeast University, Nanjing 211100, People's Republic of China\\
$^{56}$ State Key Laboratory of Particle Detection and Electronics, Beijing 100049, Hefei 230026, People's Republic of China\\
% ...
$^{69}$ University of Science and Technology of China, Hefei 230026, People's Republic of China\\
% ...
$^{72}$ University of Turin and INFN, (A)University of Turin, I-10125, Turin, Italy; (B)University of Eastern Piedmont, I-15121, Alessandria, Italy; (C)INFN, I-10125, Turin, Italy\\
$^{73}$ Uppsala University, Box 516, SE-75120 Uppsala, Sweden\\
% ...
$^{b}$ Also at the Novosibirsk State University, Novosibirsk, 630090, Russia\\
% ...
}
}

\begin{abstract}
Using $(1.0087\pm0.0044)\times10^{10}$ $J/\psi$ events collected by the BESIII detector at the BEPCII collider, we search for the lepton number violation decay $\phi \to \pi^+ \pi^+ e^- e^-$ via $J/\psi\to \phi\eta$.
No signal is found and the upper limit on the branching fraction of $\phi \to \pi^+ \pi^+ e^- e^-$ is set to be $9.7\times10^{-6}$ at the 90\% confidence level.
\end{abstract}

\maketitle
\section{Introduction}

In the Standard Model (SM) of particle physics, due to the absence of a right-handed neutrino component and the requirements of $SU(2)_L$ gauge invariance and renormalizability, neutrinos are postulated to be massless. However, the observations of neutrino oscillation~\cite{sm1,sm2,sm3,sm4} have shown that neutrinos have tiny masses, which provides evidence for physics beyond the SM. Furthermore, the observed matter-antimatter asymmetry \cite{matter} in the universe composes a serious challenge to our understanding of nature. The Big Bang theory~\cite{ref::big bang}, the prevailing cosmological model for the evolution of the universe, predicts exactly equal numbers of baryons and antibaryons in the dawn epoch. However, the observed baryon number (BN) exceeds the number of anti-baryons by a very large ratio, currently estimated at $10^9 - 10^{10}$~\cite{bn}. To give a reasonable interpretation of the baryon-antibaryon asymmetry, Sakharov proposed three principles~\cite{sa}, the first of which is that BN conservation must be violated. Many theories~\cite{Weinberg,AZee} believe that if there is BN violation, there will also be lepton number violation (LNV).

In 1937, Majorana solved the relativistic wave equation for electrons proposed by Dirac in a different way, deriving a new solution which can describe a kind of neutral fermions with masses~\cite{majorana}. As a matter of fact, neutrino masses in the extended SM can be constructed in different theoretical ways, which can lead to Dirac neutrino or Majorana neutrino. If neutrinos are Dirac particles, the neutrino and anti-neutrino are different particles. On the contrary, if they are Majorana particles, the neutrino and anti-neutrino are the same particle. However, no experiment gives us information on whether neutrinos are Dirac or Majorana particles so far~\cite{ref::pdg}. The nature of neutrinos remains an open question.

Meanwhile, in the SM, lepton number (LN) and BN are accidental global symmetries that could be violated by non-perturbative weak effects~\cite{Dur}. If the neutrino is a Majorana particle, it can be manifested by the violation of the total lepton number conservation by two units $\Delta L = 2$.
Therefore, this kind of LNV is widely viewed as the cleanest test of the Majorana neutrino.

In 1939, based on the Majorana theory, Furry~\cite{furry} studied the neutrinoless double beta ($0\nu2\beta$) decay~\cite{beta,CDEX:2023owy,Cropper:2023yrm,Arnquist:2023bhr,Zhao:2023vyh,NEXT:2023daz}, which is an LNV process.  Since then, many theories about the LNV processes have been proposed, and several experiments have been carried out to explore the LNV processes.

In recent years, many collider experiments, such as LHCb~\cite{lhc}, CMS~\cite{cms}, BaBar~\cite{babar}, ATLAS~\cite{atlas}, CLEO~\cite{cleo}, FOCUS~\cite{focus} and BESIII~\cite{bes3 neutrino}, searched for LNV processes. However, no significant evidence of a possible LNV effect is observed yet.

In the $\tau$-charm energy region, we can use the huge amount of $J/\psi$ events collected by the BESIII experiment to search for various possible LNV processes.  Considering that the decay products of $J/\psi$ particles contain a large number of intermediate particles, we can also look for LNV processes in the decays of $\phi$, $\omega$, $K^0$ and $\eta/\eta'$ mesons.

\begin{figure}[h]
\centering
\includegraphics[height=3cm,width=8cm]{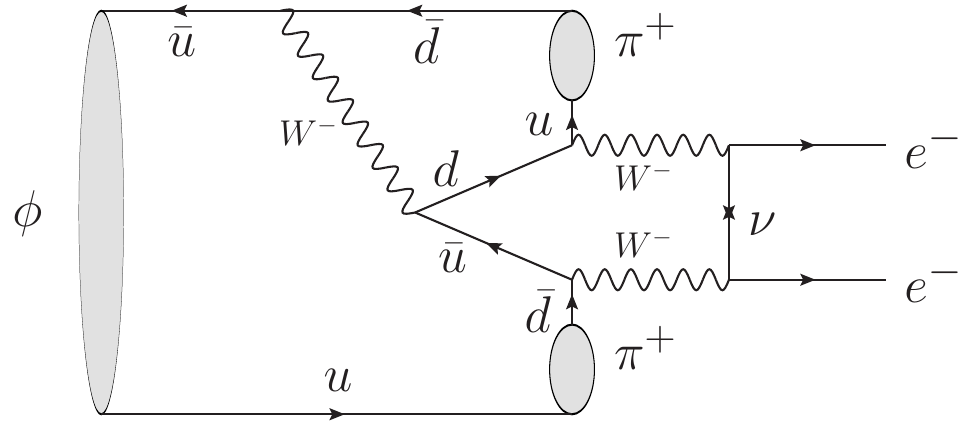}
\caption{One possible Feynman diagram of the decay $\phi\to\pi^+\pi^+e^-e^-$.}
\label{fig::feynman}
\end{figure}

As an example, Fig.~\ref{fig::feynman} shows one possible Feynman diagram for the $\phi\to\pi^+ \pi^+ e^- e^-$ decay in the scenario of Majorana neutrino, where the processes would be extremely suppressed by several loops. However, if any hint is observed with the currently available data samples, this would  definitely be a signal for new physics.

In this paper, we analyze $(1.0087\pm0.0044)\times10^{10}$ $J/\psi$ data sample collected with the BESIII detector~\cite{ref::BESCol} operating at the BEPCII storage ring~\cite{Yu:IPAC2016-TUYA01} to search for the SM forbidden LNV decay of $\phi\to\pi^+\pi^+e^-e^-$. The charge conjugation is implied throughout this paper.

\section{BESIII detector and Monte Carlo simulation}

The BESIII detector~\cite{ref::BESCol} records symmetric $e^+e^-$ collisions
provided by the BEPCII storage ring~\cite{Yu:IPAC2016-TUYA01}, which operates in $\sqrt{s}$ from 2.00 to 4.99~GeV, with a peak luminosity of $1\times10^{33}$~cm$^{-2}$s$^{-1}$ achieved at $\sqrt{s}$ = 3.77~GeV. BESIII has collected large
data samples in this energy region~\cite{Ablikim:2019hff}. The cylindrical core
of the BESIII detector covers 93\% of the full solid angle and consists of a
helium-based multilayer drift chamber~(MDC), a plastic scintillator
time-of-flight system~(TOF), and a CsI(Tl) electromagnetic calorimeter~(EMC),
which are all enclosed in a superconducting solenoidal magnet providing a 1.0~T
magnetic field. The magnetic field was 0.9~T in 2012, which affects 11\% of the total $J/\psi$ data.. The solenoid is supported by an octagonal flux-return yoke with
resistive plate counter muon identification modules interleaved with steel.
The charged-particle momentum resolution at $1~{\rm GeV}/c$ is $0.5\%$, and the specific ionization energy loss (d$E$/d$x$) resolution is $6\%$ for electrons from Bhabha scattering.
The EMC measures photon energies with a resolution of $2.5\%$ ($5\%$) at
$1$~GeV in the barrel (end cap) region. The time resolution in the TOF barrel
region is 68~ps, while that in the end cap region is 110~ps. The end cap TOF
system was upgraded in 2015 using multigap resistive plate chamber technology,
providing a time resolution of 60~ps,
which benefits 87\% of the data used in this analysis~\cite{etof}.

Simulated Monte Carlo~(MC) samples produced with {\sc geant4}-based~\cite{ref::geant4}
software, which includes the geometric description of the BESIII detector~\cite{ref::detector} and
the detector response, are used to determine the detection efficiency and to
estimate the background contributions. The simulation includes the beam-energy
spread and initial-state radiation~(ISR) in the $e^+e^-$ annihilation modeled
with the generator {\sc kkmc}~\cite{ref::kkmc}.
The inclusive MC simulation sample includes both the production of the $J/\psi$ resonance and the
continuum processes incorporated in {\sc kkmc}~\cite{ref::kkmc}.
The known decay modes are modeled with {\sc evtgen}~\cite{ref::evtgen} using world averaged branching fraction
values~\cite{ref::pdg}, and the remaining unknown decays from the charmonium states
with {\sc lundcharm}~\cite{ref::ISR-lundcharm}. Final-state radiation from charged
final-state particles is incorporated with {\sc photos}~\cite{PHOTOS}.

\section{Data analysis}
\subsection{Method}
In this analysis, we search for the decay $\phi\to\pi^+\pi^+e^-e^-$ via $J/\psi\to\phi\eta, \eta\to\gamma\gamma$. In order to avoid the large uncertainty from  $\mathcal{B}(J/\psi\to\phi\eta)$~\cite{ref::pdg}, which is about 11\%, we measure the branching fraction of the signal decay $\phi\to\pi^+\pi^+e^-e^-$ relative to that of the reference channel $\phi\to K^+K^-$ via $J/\psi\to\phi\eta$.

The branching fractions of $\phi\to \pi^+\pi^+ e^-e^-$ and $\phi\to K^+K^-$ can be written as
\begin{equation}
\begin{split}
\mathcal{B}(\phi \to\pi^+\pi^+e^-e^-)=\frac{N_{\pi^+\pi^+e^-e^-}^{\rm{net}}/\varepsilon_{\pi^+\pi^+e^-e^-}}{N^{\rm{tot}}\times \mathcal{B}(J/\psi\to\phi\eta)\times\mathcal{B}(\eta\to\gamma\gamma)},
\end{split}
\end{equation}
and
\begin{equation}
\small
\mathcal{B}(\phi \to K^+K^-)=
\frac{N_{K^+K^-}^{\rm{net}}/\varepsilon_{K^+K^-}}{N^{\rm{tot}}\times \mathcal{B}(J/\psi\to\phi\eta)\times\mathcal{B}(\eta\to\gamma\gamma)},
\end{equation}
respectively, where $N_{\pi^+\pi^+e^-e^-}^{\rm{net}}$ and $N_{ K^+K^-}^{\rm{net}}$ are the signal yields, $N^{\rm{tot}}$ is the  total number of $J/\psi$ events,  $\varepsilon_{\pi^+\pi^+e^-e^-}$ and $\varepsilon_{K^+K^-}$ are the detection efficiencies for the $J/\psi\to\eta\phi,\eta\to\gamma\gamma,\phi\to\pi^+\pi^+e^-e^-$ and $J/\psi\to\eta\phi,\eta\to\gamma\gamma,\phi\to K^+K^-$, respectively.  $\mathcal{B}(J/\psi\to\phi\eta)$ and $\mathcal{B}(\eta\to\gamma\gamma)$ are the branching fractions of $J/\psi\to\phi\eta$ and $\eta\to\gamma\gamma$. With the above two equations, the branching fraction of $\phi\to \pi^+\pi^+ e^-e^-$ can be determined by
\begin{equation}
\begin{split}
\mathcal{B}(\phi \to\pi^+\pi^+ e^-e^-)=\mathcal{B}(\phi\to K^+K^-)\\ \times\frac{N_{\pi^+\pi^+e^-e^-}^{\rm{net}}/\varepsilon_{\pi^+\pi^+e^-e^-}}{N_{K^+K^-}^{\rm{net}}/\varepsilon_{K^+K^-}},\label{bf}
\end{split}
\end{equation}
where $\mathcal{B}(\phi\to K^+K^-)=(49.2\pm0.5)\%$~\cite{ref::pdg}. The uncertainty of the input $\mathcal{B}(\phi\to K^+K^-)$ is only 1.0\%,  thus the total systematic uncertainty can be reduced significantly.

\subsection{Analysis of $\phi\to K^+K^-$}
The reference decay $J/\psi\to \phi\eta$ is reconstructed with $\eta\to\gamma\gamma$ and $\phi\to K^+K^-$. In each event, at least two charged tracks and two neutral candidates are required.

Charged tracks detected in the MDC are required to be within a polar angle ($\theta$) range of $|\rm{cos\theta}|<0.93$, where $\theta$ is defined with respect to the $z$-axis, which is the symmetry axis of the MDC. The distance of closest approach to the interaction point (IP) must be less than 10\,cm along the $z$-axis, $|V_{z}|$,  and less than 1\,cm
in the transverse plane, $|V_{xy}|$. Events with exactly two good charged tracks with zero net charge are kept for further analysis.
For charged particle identification (PID), we use a combination of the $dE/dx$ in the MDC, and the time of flight in the TOF to calculate the Confidence Level (CL) for pion and kaon hypotheses ($CL_{\pi}$ and $CL_K$). For kaon candidates, they are required to satisfy $CL_K>0.001$ and $CL_K>CL_{\pi}$ to avoid contamination from pions and to suppress background.

The photon candidates are selected from isolated EMC clusters. To suppress the electronics noise and beam related background, the clusters are required to start within $700\ \mathrm{ns}$ after the event start time and fall outside a cone angle of $20^\circ$ around the nearest extrapolated good charged track.
The minimum energy of each EMC cluster is required to be greater than 25 MeV in the barrel region ($|\cos\theta|<0.80$) or 50 MeV in the end-cap regions ($0.86<|\cos\theta|<0.92$).
The $\eta$ candidate is reconstructed by $\eta \to \gamma \gamma$, where the invariant mass of the $\gamma \gamma$ pair is required to satisfy 0.45 GeV/${c}^2$$ < M_{\gamma \gamma} <$ 0.65 GeV/${c}^2$.
To reduce backgrounds and improve mass resolution, a kinematic fit is performed by constraining the total four momentum (4C) to that of the initial $e^+e^-$ beams, under the hypothesis of $e^+e^-\to K^+K^-\gamma\gamma$. All good photons are looped over together with the two tracks in the kinematic fit, and candidate with  the least $\chi^2$ is retained for further analysis.

\begin{figure}[h!]
\centering
\includegraphics[width=0.48\textwidth]{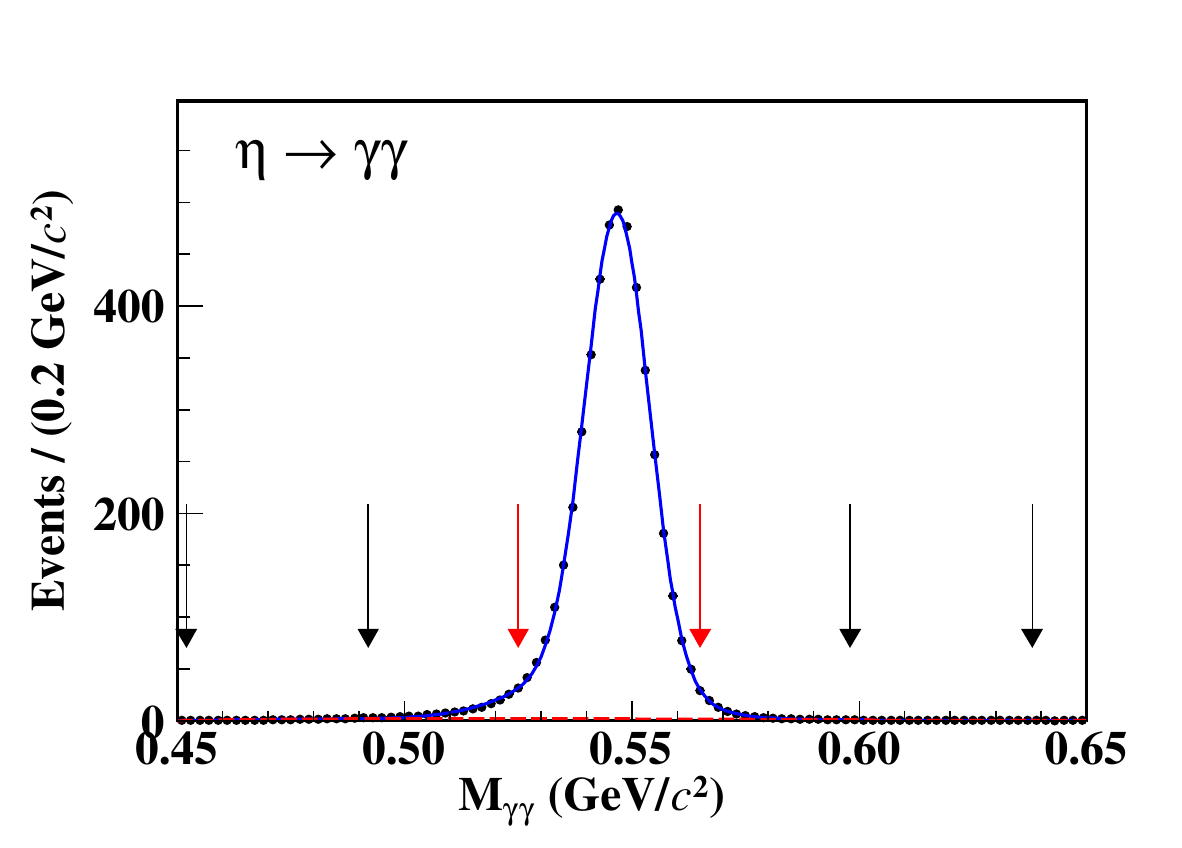}
\caption{Fit to the $M_{\gamma\gamma}$ distribution, where the black points represent the signal MC candidates, the blue curve is the fit result, the red dashed line is the second-order polynomial background, the red arrows show the signal region and the black arrows show the sideband region.}
\label{fig::gamgam}
\end{figure}

\begin{figure}[h!]
\centering
\includegraphics[width=0.47\textwidth]{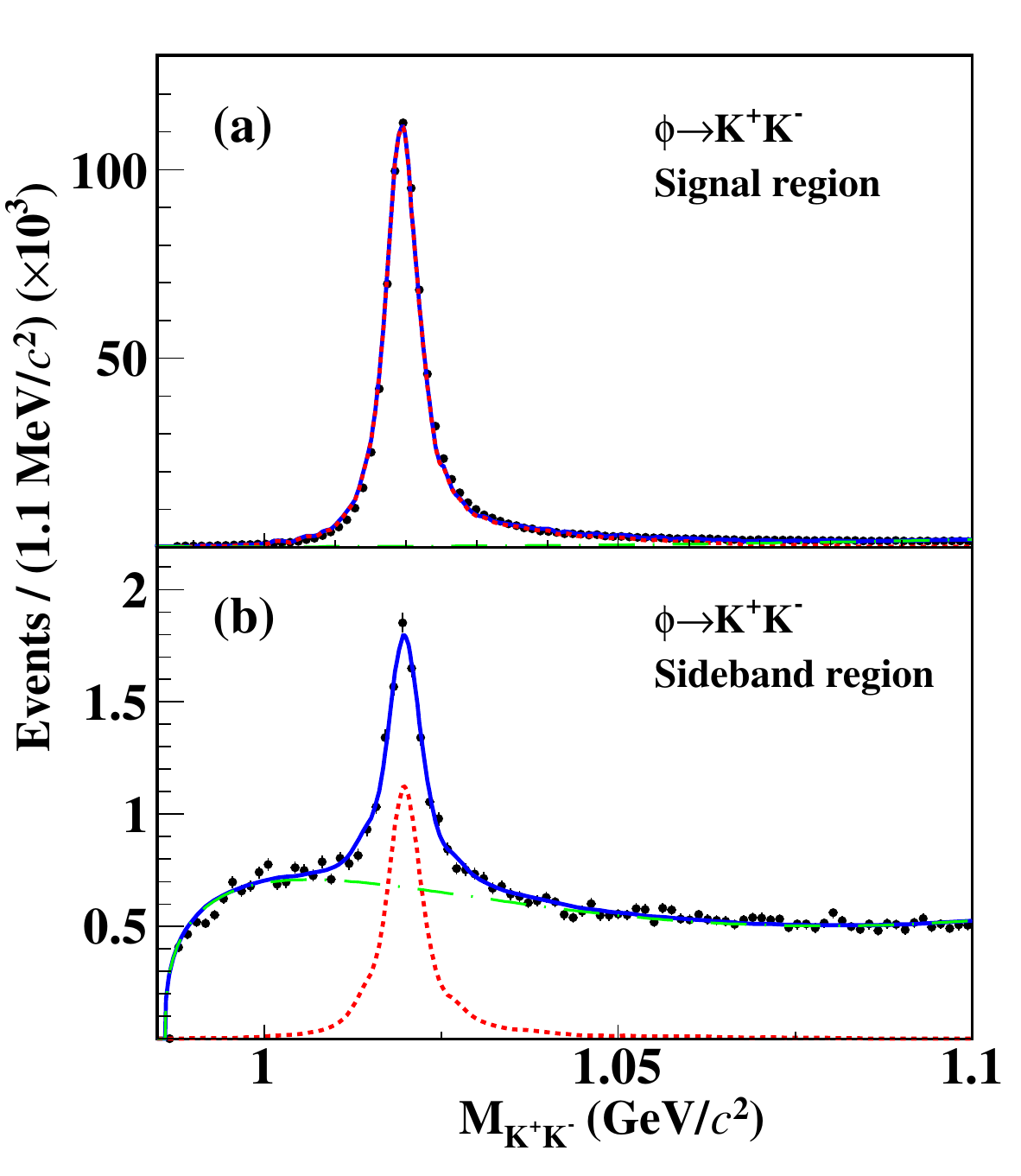}
\caption{Invariant mass $M_{K^+K^-}$ distributions of $J/\psi  \to \phi \eta$ candidates in the (a) signal and (b) sideband region of $\eta$, with fit results overlaid, where the black points represent data, the blue curves are the fit results, the green dash-dotted curves are the fitted background shapes, the red dotted curves are the signal shapes.}
\label{fig::fit}
\end{figure}

To obtain the signal yield of $\phi\to K^+K^-$, we fit the $M_{K^+K^-}$ distribution of the accepted candidate events in the $\eta$ signal region ([0.525,0.565]~GeV/${c}^2$) and in the  $\eta$ sideband region ([0.452,0.492]~GeV/${c}^2$ \& [0.598,0.638]~GeV/${c}^2$). The signal region is determined by $[\mu-3\sigma, \mu+3\sigma]$; the sideband region is outside the signal region and the distance between the two intervals is 5$\sigma$, where $\mu$ and $\sigma$ are mean and standard deviation obtained from a Gaussian fit to the  $M_{\gamma\gamma}$ distribution as shown in Fig.~\ref{fig::gamgam}. In the fits to the $M_{K^+K^-}$ distributions in the $\eta$ signal region and sideband region, the signal shape is described by a truth-matched MC shape convolved with a double Gaussian function and the background shape is described by an inverted ARGUS function~\cite{argus} multiplied by a fourth-order polynomial function. The parameters of the double Gaussian function and the background model are floating in the fits.
With the numbers of $N_{\rm signal}$ and $N_{\rm sideband}$ obtained from the fits in Fig.~\ref{fig::fit}, the net number of the $\phi\to K^+K^-$ candidate events is calculated by
\begin{equation}
N^{\rm net}_{K^+K^-}=N_{\rm signal}-\frac{1}{2}\times N_{\rm sideband}=822665\pm1149.
\end{equation}

To determine the detection efficiency, the $J/\psi\to \phi\eta~(\phi\to K^+K^-, \eta\to\gamma\gamma)$ decays are simulated, where the decays of $J/\psi\to\phi\eta$, $\phi\to K^+K^-$ and $\eta\to\gamma\gamma$ are modeled by a helicity amplitude generator {\sc HELAMP}, VSS model (decay of a vector particle to a pair of scalars), and a phase space ({\sc PHSP}) generator, respectively~\cite{Generator}. After applying all the selection criteria, we fit the invariant mass of the $K^+K^-$ ($M_{K^+K^-}$) for the survived signal MC events in the signal region and sideband region, respectively. In the fits, the signal shape is modeled by a truth-matched MC shape convolved with a double Gaussian function and the background is described by a second-order polynomial function. The number of observed signal events is obtained to be $N^{\rm MC}_{\rm obs}=N_{\phi\to K^+K^-}^{\rm signal}-\frac{1}{2}\times N_{\phi\to K^+K^-}^{\rm sideband}$, where $N_{\phi\to K^+K^-}^{\rm signal}$ and $N_{\phi\to K^+K^-}^{\rm sideband}$ are the observed numbers in the $\eta$ signal region and sideband region, respectively. Dividing it by the number of total signal MC events $N^{\rm MC}_{\rm total}$, the efficiency of detecting the decays of $\phi\to K^+K^-$ is
\begin{equation}
\varepsilon_{KK} = \frac{N^{\rm MC}_{\rm obs}}{N^{\rm MC}_{\rm total}}=\frac{471297}{1000000}=47.1\%.
\end{equation}

\subsection{Analysis of $\phi\to\pi^+\pi^+ e^-e^-$}

The LNV decay of $\phi \to \pi^+\pi^+e^-e^-$ is searched through $J/\psi\to \phi\eta$, where the $\eta$ candidate is reconstructed through $\eta\to\gamma\gamma$. In each event, at least four charged tracks and two neutral candidates are required.

The good charged tracks are selected with the same criteria as the reference mode. For charged PID, we use a combination of the $dE/dx$ in the MDC, the time of flight in the TOF, and the energy and shape of clusters in the EMC to calculate the CL for electron, pion, and kaon hypotheses ($CL_e$, $CL_{\pi}$, and $CL_K$).
The electron candidates are required to satisfy $CL_e>0.001$ and $CL_e/(CL_e+CL_K+LC_{\pi})>0.8$. Furthermore, to suppress background from pions, electrons must satisfy an additional requirement $E/p > 0.8$ for tracks with $p_e \geq$ 0.5  GeV/$c$, where $E$ and $p$ represent the energy deposited in the EMC and the momentum reconstructed in the MDC, respectively. Pion candidates are required to satisfy $CL_\pi>0$ and $CL_\pi>CL_K$. The good photons are selected in the same way as in the reference mode.

To reduce backgrounds and improve the mass resolution, a kinematic fit is performed by constraining the total four momentum (4C) to that of the initial $e^+e^-$ beams. All the good photons are looped over together with the four tracks in the kinematic fit. The $\chi^2$ of the 4C kinematic fit is required to be less than 30, which is determined through the Punzi significance method~\cite{punzi} with the formula $\frac{\varepsilon}{1.5+\sqrt{B}}$, where $\varepsilon$ is the detection efficiency and $B$ is the number of background events from the inclusive MC sample. If there is more than one combination, the combination with the minimum $\chi^2$ is retained.

To suppress contamination of the final states with four charged tracks from mis-identification, we perform 4C kinematic fits under six different hypotheses of $J/\psi \to \pi^+\pi^+e^-e^-\gamma\gamma$, $K^+K^-K^+K^-\gamma\gamma$, $K^+K^- p\bar p\gamma\gamma$, $K^+K^-\pi^+\pi^-\gamma\gamma$, $\pi^+\pi^-\pi^+\pi^-\gamma\gamma$, and $\pi^+\pi^-p\bar p\gamma\gamma$.
If the kinematic fit for the $\pi^+\pi^+e^-e^-\gamma\gamma$ hypotheses is successful and gives the minimum $\chi^2$ among these six assignments, the event is then accepted for further analysis.

To further suppress possible background from $\gamma-$conversion, the opening angle $\theta_{\pi e} $ between any pions (possibly mis-identified from real positrons) and electrons are required to be greater than $8^\circ$.

Based on a fit with a double Gaussian function and a Chebychev polynomial to model the signal and background shapes of the simulated $M_{\pi^+\pi^+ e^-e^-}$ and $M_{\gamma\gamma}$ distributions, the signal region is determined to be $[0.99, 1.04]$ GeV/${c}^2$ for $M_{\pi^+\pi^+ e^-e^-}$ and $[0.52, 0.57]$ GeV/${c}^2$ for $M_{\gamma\gamma}$. This corresponds to a range of $\pm3$ times the mass resolution around their known masses~\cite{ref::pdg}.
The detection efficiency is determined to be 4.40\% with simulated $J/\psi\to\phi\eta \to (\pi^+\pi^+e^-e^-)(\gamma\gamma)$ events, where the $J/\psi$ decay is modeled by a helicity amplitude generator {\sc HELAMP}~\cite{Generator} and the $\phi$/$\eta$ decays are modeled by a phase space ({\sc PHSP}) generator.

\begin{figure}[htbp]
\centering
\includegraphics[width=0.48\textwidth]{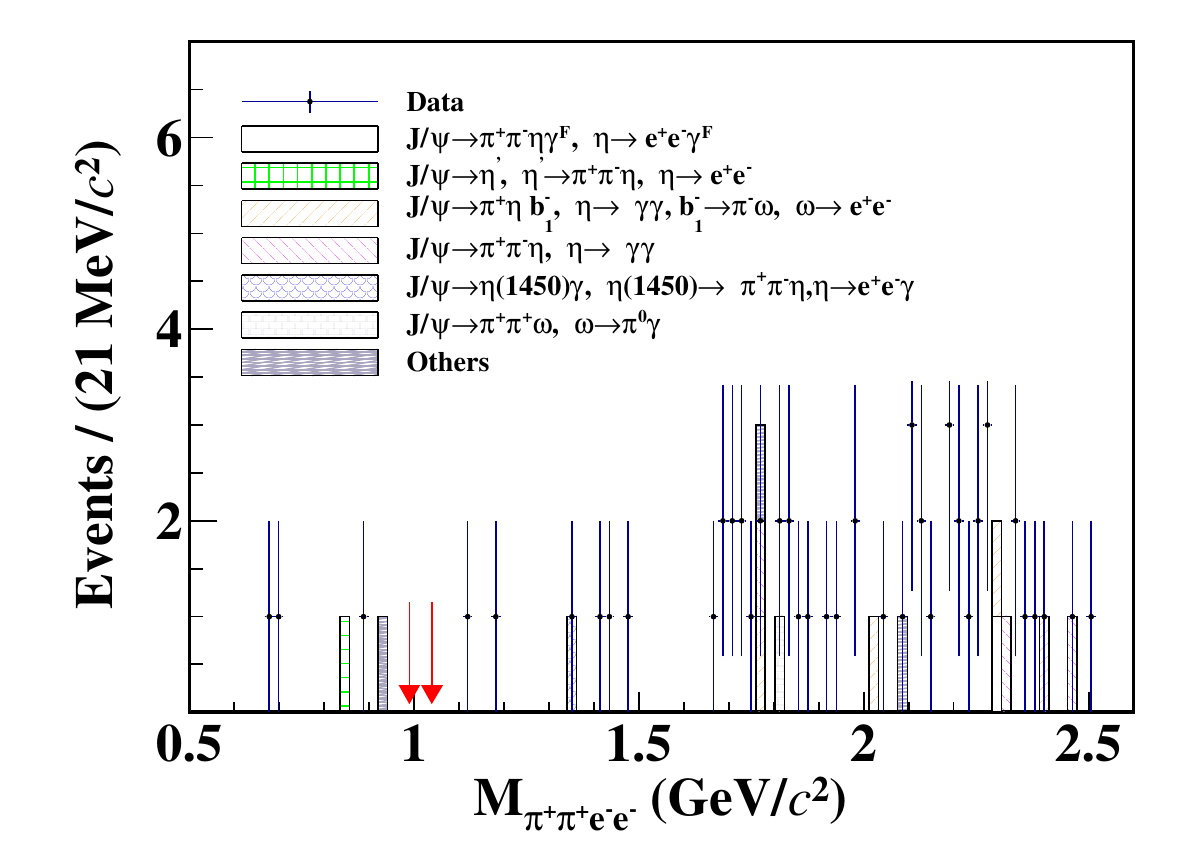}
\caption{
The distribution of $M_{\pi^+\pi^+ e^-e^-}$ for the events in the range of $M_{\gamma\gamma}\in(0.52,0.57)$ GeV/$c^2$, where the points with error bars are data, the histogram in different styles represent different sources of background modes shown in legend. The red arrows show the signal region.}
\label{fig::bkg}
\end{figure}

With an event type analysis tool, TopoAna~\cite{topoana}, the backgrounds from $J/\psi$ decays are investigated using an inclusive MC simulation sample, which has the same size as the $J/\psi$ data sample. Only 39 events from 14 different decay channels remain and all of them are located far away from the signal region.
The distribution of $M_{\pi^+\pi^+ e^-e^-}$ for the background events from the inclusive MC simulation sample is shown in Fig.~\ref{fig::bkg}, where the red arrows show the signal region.
No background event is found near the signal region.

To avoid the influence of statistical fluctuation, large exclusive MC simulation samples for the three main background channels, (1) $J/\psi \to \pi^{+}\pi^-\eta\gamma^F,\eta \to e^{+}e^-\gamma^F$ ($\gamma^F$ is the $\gamma$ from final state radiation), (2) $J/\psi \to\eta^{\prime}, \eta^{\prime}\to\pi^{+}\pi^-\eta,\eta \to e^+e^-$, (3) $J/\psi \to \pi^{+}\eta b_{1}^{-},\eta \to \gamma \gamma, b_{1}^{-}\to \pi^{-}\omega,\omega \to e^{+}e^-$, are produced. Furthermore, possible background from other $\phi$ decays, such as $J/\psi \to \phi \eta, \phi\to e^{+}e^-\eta, \eta \to \gamma \gamma$, is also checked. No background event is found near the signal region.

\begin{figure}[h!]
\centering
\includegraphics[width=0.48\textwidth]{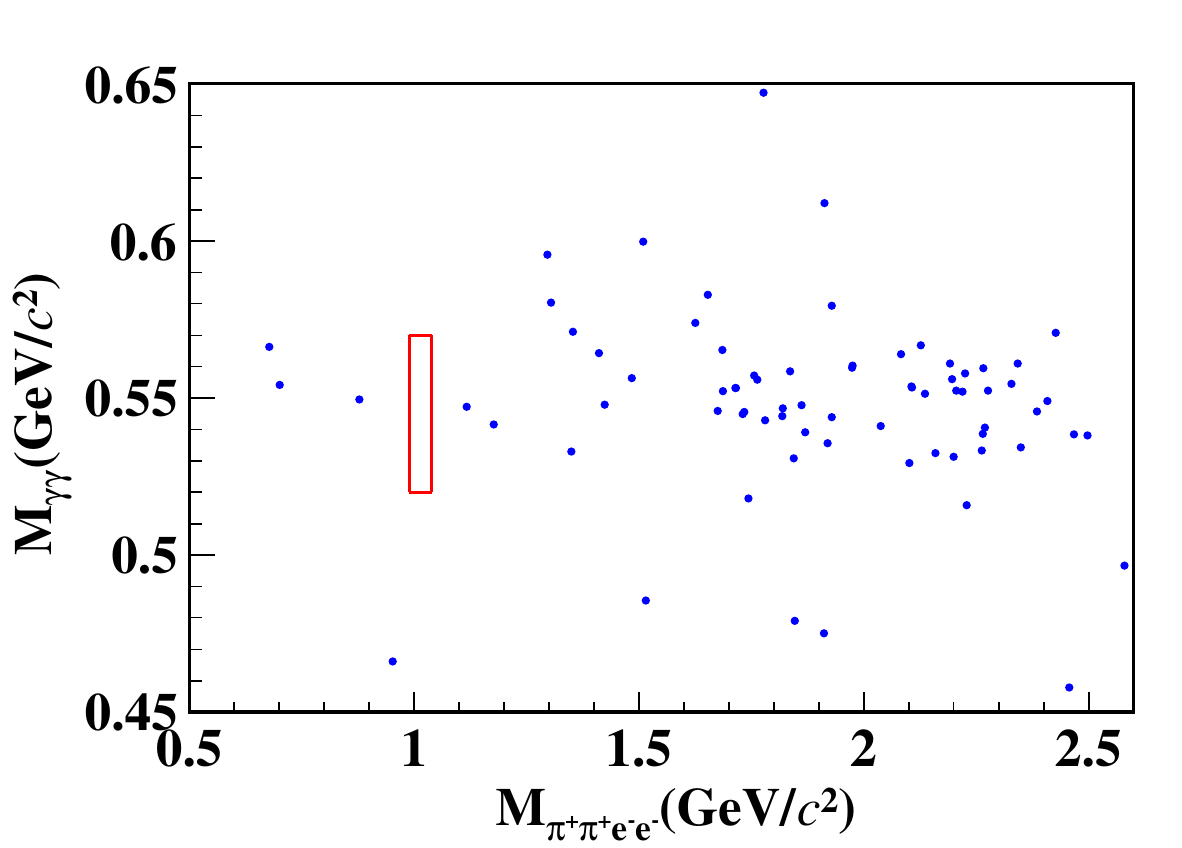}
\caption{The distribution of $M_{\gamma\gamma}$ versus $M_{\pi^+ \pi^+ e^-e^-}$ of the accepted $\phi \to \pi^+ \pi^+ e^-e^-$ candidate events in data. The red box indicates the signal region defined as  $[0.99, 1.04]$ GeV/${c}^2$ for $M_{\pi^+\pi^+ e^-e^-}$ and $[0.52, 0.57]$ GeV/${c}^2$ for $M_{\gamma\gamma}$.}
\label{fig::data}
\end{figure}

Figure~\ref{fig::data} shows the two dimensional distribution of $M_{\gamma\gamma}$ versus $M_{\pi^+ \pi^+ e^-e^-}$ of the accepted $\phi \to \pi^+ \pi^+ e^-e^-$ candidate events in data. There is no event in the signal region. The signal yield ($N^{\rm sig}$) and the background yield ($N^{\rm bkg}$) are determined to be 0.

\section{Systematic Uncertainty}
The sources of systematic uncertainties for the product branching fractions include MDC tracking, charged PID, 4C kinematic fit, $\chi^2$ requirement, $\theta_{\pi e}$ requirement, signal window, fitting procedure, MC modeling, $N^{\rm net}_{K^+K^-}$ determination, and $\mathcal{B}(\phi\to K^+K^-)$. All the systematic  uncertainties are summarized in Table \ref{tab::sys}, and the total uncertainty is obtained by adding the individual components in quadrature.

\begin{table}[h!]
\centering
\caption{Relative systematic uncertainties in the branching measurement.}
\begin{tabular}{lc}
\hline
\hline
Source & Uncertainty (\%)\\ \hline
MDC tracking                                        & 2.6\\
PID                                                 & 6.2\\
4C kinematic fit for $\phi\to K^+K^-$               & 0.2\\
4C kinematic fit for $\phi\to \pi^+\pi^-e^+e^-$     & 2.3\\
$\theta_{\pi e}$ selection requirement            & 4.1\\
Signal window                   & 0.2\\
Yield of $\phi\to K^+K^-$    & 1.1\\
MC modeling                     & 1.9\\
$\mathcal{B}(\phi\to K^+K^-)$   & 1.0\\
\hline
Total                           & 8.6\\
\hline
\hline
\end{tabular}
\label{tab::sys}
\end{table}

According to Eq.~\ref{bf}, the systematic uncertainties of photon detection and quoted branching fractions ($\mathcal{B}(J/\psi\to\phi\eta)$ and $\mathcal{B}(\eta\to\gamma\gamma)$) are cancelled.

The uncertainties on tracking efficiency are estimated with the control samples $J/\psi \to\pi^+\pi^-\pi^0$, $J/\psi\to e^+e^-(\gamma_{{\rm FSR}})$  ($\gamma_{{\rm FSR}}$ is the FSR photon) and $J/\psi\to\pi^0K^+K^-$, and are determined to be 0.3\% per pion, 0.7\% per electron and 0.3\% per kaon, respectively. Similarly, the uncertainties of PID are 1.0\%, 1.0\%, and 1.1\% for each charged electron, pion, and kaon, respectively. After adding the systematic uncertainties of each track linearly, the total systematic uncertainties on the tracking efficiency and PID efficiency are obtained to be 2.6\% and 6.0\%, respectively.

The systematic uncertainty due to the 4C kinematic fit for $J/\psi\to\phi\eta\to K^+K^-\eta~(\eta\to\gamma\gamma)$ is studied by using the control sample of $J/\psi\to K^+K^-\pi^0~(\pi^0\to\gamma\gamma)$ decay mode. The corresponding uncertainty is estimated to be 0.2\% by comparing the difference of efficiencies between data and MC simulation. Similarly, the systematic uncertainty due to the 4C kinematic fit and $\chi^2<30$ for $J/\psi\to \phi\eta \to \pi^+\pi^-e^+e^-\eta~(\eta \to \gamma\gamma)$ can be studied by using control sample of $J/\psi \to \pi^+\pi^-\pi^+\pi^-\eta~(\eta \to \gamma\gamma)$ decay mode. The corresponding uncertainty is assigned to be 2.3\%.

The uncertainty of the $\theta_{\pi e}$ requirement is estimated by varying the optimized requirement $\theta_{\pi e}>8^\circ$ with alternative  $\theta_{\pi e}$ requirements, i.e. $\theta_{\pi e}>3^\circ$, $\theta_{\pi e}>4^\circ$, ..., $\theta_{\pi e}>12^\circ$, $\theta_{\pi e}>13^\circ$. The standard deviation on the detection efficiency, 4.1\%, is taken as the corresponding systematic uncertainty.

To investigate the systematic uncertainty due to the choice of $\eta$ signal window, we use different signal window ranges, such as $\pm 3.1\sigma$, $\pm 3.2\sigma$, $\pm 2.8\sigma$, etc. The standard deviation on the detection efficiency of 0.2\% is taken as the uncertainty.

The systematic uncertainty of the yield of the reference decay $J/\psi\to\phi\eta, \phi\to K^+K^-$ includes the fit range, the signal shape, and the background shape. The uncertainty due to the fit range of $M_{KK}$ is estimated by changing the fit from (0.99,1.10) GeV/${c}^2$ to (0.99,1.09) GeV/${c}^2$ and (0.98,1.10) GeV/${c}^2$.
The uncertainty due to the background shape is estimated by changing the second-order  polynomial function to a first-order polynomial function.
To estimate the systematic uncertainty due to signal shape, we use alternative signal shapes, an MC shape convolved with a Gaussian function. The relative difference between the signal yield and the detection efficiency is taken as the corresponding systematic uncertainty. As a result, the systematic uncertainties are 1.0\%, 0.1\%, and 0.3\% for fit range, background shape, and signal shape, respectively. After adding them in quadrature, the total systematic uncertainty associated with the fit procedure is obtained to be 1.1\%.

In order to estimate the uncertainty related to the MC simulation model, the Majorana neutrino $\nu_N$ is assumed to decay into $\pi^+e^-$, and it has a spin of 1/2, a charge of zero and a width of zero. However, the $\nu_N$ mass remains unknown and can range from the $\pi e$ mass threshold to the largest available phase space of $\phi$ decay, which needs to satisfy $m_e+m_\pi \leq m_{\nu_N} \leq \frac{m_\phi}{2}$.  We divide the mass range (0.150,0.50) GeV into 14 equidistant intervals, with a step of $0.025$ GeV, i.e., 0.175 GeV, 0.200 GeV, ..., 0.500 GeV. The detection efficiency is averaged to be $(14.30 \pm 0.16)\%$. The difference between this value and the detection efficiency obtained by the {\sc PHSP} model of 1.9\% is taken as the associated systematic uncertainty.

The uncertainty of the quoted branching fraction $\mathcal{B}(\phi\to K^+K^-)$ is 1.0\%.
\section{Result}
Because no event is observed in the signal region, the upper limit on the signal yield $N_{\pi^+\pi^+ e^-e^-}^{\rm up}$ is estimated to be $37.1$ at the 90\% CL by utilizing a frequentist method \cite{ref::TROLKE} with unbounded profile likelihood treatment of systematic uncertainties, where the background fluctuation is assumed to follow a Poisson distribution, the detection efficiency ($\varepsilon^{\rm sig}=4.40\%$) is assumed to follow a Gaussian distribution, and the systematic uncertainty ($\Delta_{\rm sys}=8.6\%$) is considered as the standard deviation of the efficiency.

The upper limit on the branching fraction of $\phi\to \pi^+\pi^+ e^-e^-$ is determined by
\begin{equation}
\mathcal{B}(\phi\to \pi^+\pi^+ e^-e^-) < \mathcal{B}(\phi\to K^+K^-)\times \frac{N_{\pi^+\pi^+ e^-e^-}^{\rm up} }{N_{K^+K^-}^{\rm net}/ \varepsilon_{K^+K^-}},
\nonumber
\end{equation}
where $\varepsilon_{K^+K^-}=47.1\%$, $N_{K^+K^-}^{\rm net}=822665 \pm 1149$, $\mathcal{B}(\phi\to K^+K^-)=(49.2 \pm 0.5)\%$ \cite{ref::pdg} and $N_{\pi^+\pi^+ e^-e^-}^{\rm up}=37.1$. Thus, the upper limit on the branching fraction is set to be
\begin{equation}
\mathcal{B}(\phi\to\pi^+\pi^+ e^-e^-)<3.7\times 10^{-6}.
\nonumber
\end{equation}

\section{Summary}
In summary, by analyzing $(1.0087\pm0.0044)\times10^{10}$ $J/\psi$ events collected with the BESIII detector at the BEPCII collider, we search for the LNV decay $\phi \to \pi^+ \pi^+ e^-e^-$ via $J/\psi \to \phi \eta$ for the first time.
No signal event has been observed and the upper limit on the branching fraction of this decay is set to be $3.7\times 10^{-6}$ at the $90$\% CL. This is the first constraint of LNV signal in $\phi$ meson decays. Our result improves the experimental knowledge of neutrinoless double beta decay for the hadrons composed of second generation quarks.

\section{Acknowledgements}
The BESIII Collaboration thanks the staff of BEPCII and the IHEP computing center for their strong support. This work is supported in part by National Key R\&D Program of China under Contracts Nos. 2020YFA0406300, 2020YFA0406400; National Natural Science Foundation of China (NSFC) under Contracts Nos. 12035009, 11875170, 11635010, 11735014, 11835012, 11935015, 11935016, 11935018, 11961141012, 12022510, 12025502, 12035013, 12061131003, 12192260, 12192261, 12192262, 12192263, 12192264, 12192265; the Chinese Academy of Sciences (CAS) Large-Scale Scientific Facility Program; the CAS Center for Excellence in Particle Physics (CCEPP); Joint Large-Scale Scientific Facility Funds of the NSFC and CAS under Contract No. U1832207; CAS Key Research Program of Frontier Sciences under Contracts Nos. QYZDJ-SSW-SLH003, QYZDJ-SSW-SLH040; 100 Talents Program of CAS; The Institute of Nuclear and Particle Physics (INPAC) and Shanghai Key Laboratory for Particle Physics and Cosmology; ERC under Contract No. 758462; European Union's Horizon 2020 research and innovation programme under Marie Sklodowska-Curie grant agreement under Contract No. 894790; German Research Foundation DFG under Contracts Nos. 443159800, 455635585, Collaborative Research Center CRC 1044, FOR5327, GRK 2149; Istituto Nazionale di Fisica Nucleare, Italy; Ministry of Development of Turkey under Contract No. DPT2006K-120470; National Research Foundation of Korea under Contract No. NRF-2022R1A2C1092335; National Science and Technology fund; National Science Research and Innovation Fund (NSRF) via the Program Management Unit for Human Resources \& Institutional Development, Research and Innovation under Contract No. B16F640076; Polish National Science Centre under Contract No. 2019/35/O/ST2/02907; Suranaree University of Technology (SUT), Thailand Science Research and Innovation (TSRI), and National Science Research and Innovation Fund (NSRF) under Contract No. 160355; The Royal Society, UK under Contract No. DH160214; The Swedish Research Council; U. S. Department of Energy under Contract No. DE-FG02-05ER41374.

\end{document}